\documentclass{article}

\usepackage[nonatbib,preprint]{neurips_data_2024}

\usepackage[utf8]{inputenc} 
\usepackage[T1]{fontenc}    
\usepackage{hyperref}       
\usepackage{url}            
\usepackage{booktabs}       
\usepackage{amsfonts}       
\usepackage{nicefrac}       
\usepackage{microtype}      
\usepackage{xcolor}         
\usepackage{graphicx}
\usepackage{caption}

\usepackage{times}
\usepackage{latexsym}
\usepackage[utf8]{inputenc}
 \usepackage{inconsolata}
 \usepackage{biblatex}
 \usepackage{comment}
\newcommand{\tool}{$\mathsf{TRUCE}$}
\title{\tool: Private Benchmarking to Prevent Contamination and Improve Comparative Evaluation of LLMs}

\author{%
Tanmay Rajore \\
\And  Nishanth Chandran \\
\And Sunayana Sitaram \\
\And Divya Gupta \\
\AND Rahul Sharma \\
\And Kashish Mittal \\
\And Manohar Swaminathan\\ 
\AND \\
Microsoft Research India \\
\texttt{tanmayrajore@gmail.com},\\
\texttt{\{nichandr,sunayana.sitaram,divya.gupta,}\\
\texttt{
rahsha,kmittal,manohar.swaminathan\}@microsoft.com} \\
}

\begin{document}

\maketitle

\begin{abstract}
Benchmarking is the de-facto standard for evaluating LLMs, due to its speed, replicability and low cost. However, recent work has pointed out that the majority of the open source benchmarks available today have been contaminated or leaked into LLMs, meaning that LLMs have access to test data during pretraining and/or fine-tuning. This raises serious concerns about the validity of benchmarking studies conducted so far and the future of evaluation using benchmarks. To solve this problem, we propose Private Benchmarking, a solution where test datasets are kept private and models are evaluated without revealing the test data to the model. We describe various scenarios (depending on the trust placed on model owners or dataset owners), and present solutions to avoid data contamination using private benchmarking. For scenarios where the model weights need to be kept private, we describe solutions from confidential computing and cryptography that can aid in private benchmarking.
We build an end-to-end system, \tool, that enables such private benchmarking showing that the overheads introduced to protect models and benchmark are negligible (in the case of confidential computing) and tractable (when cryptographic security is required). Finally, we also discuss solutions to the problem of benchmark dataset auditing, to ensure that private benchmarks are of sufficiently high quality. \end{abstract}
\section{Introduction}

Large Language Models (LLMs) have become increasingly popular due to their impressive performance in various tasks and domains, often surpassing human capabilities. However, it is crucial for LLMs to not only excel in their tasks but also be grounded in factual information and avoid generating harmful content. Therefore, evaluating and understanding the limitations and capabilities of LLMs has become an important topic of research.


LLMs are commonly evaluated through automated techniques like benchmarking, or by human interaction and feedback. Holistic evaluation of LLMs \cite{liang2022holistic} evaluates LLMs on several dimensions such as accuracy, calibration, bias and toxicity. Several popular benchmarks have been proposed such as BigBench \cite{srivastava2023beyond} to increase the task and language coverage of benchmarking. However, benchmarking has a number of limitations. For instance, the tasks and datasets provided in standard open source benchmarks may not always be applicable in real-world scenarios. Additionally, certain benchmarks may be of low quality due to inadequate or biased instructions \cite{parmar2023don} or limited training of annotators \cite{geva2019we}. Moreover, some benchmarks may contain unintended artifacts introduced by annotators' subconscious biases \cite{poliak2018hypothesis}. Nevertheless, in spite of these constraints, benchmarking remains a fast, cost-effective, and replicable means of comparing multiple models or versions of a single model. As such, benchmarking has become the de-facto standard for evaluating the capabilities of LLMs, showing superiority of one LLM over another and deciding which version of a model to ship in production.


However, the issue of test dataset contamination or leakage has raised significant concerns about the reliability of benchmarking and the validity of its outcomes. Benchmark datasets are commonly found on the internet, allowing large web crawlers that collect LLM pretraining data to potentially incorporate them into pretraining data. Another concerning issue, as pointed out by \cite{balloccu2024leak}, is that LLM researchers could have inadvertently leaked benchmark datasets to closed source models during LLM evaluation, and this data can be used for fine-tuning or prompt optimization. According to estimates based on their survey, over 4 million data points have been leaked to closed models such as ChatGPT. Creating benchmarks can be a costly process due to the human effort and expertise required for data creation and annotation. Therefore, it is imperative to prevent contamination of benchmarks.

In this paper, we introduce the notion of Private Benchmarking - a solution that keeps the benchmark dataset private from the model, while revealing only the results of the evaluation. This ensures that the model never sees the test dataset. There are several possible scenarios - the model may be open source, in which case it can be hosted locally and a private benchmark can be run on it. In case of closed models, we may trust the model owner not to use any benchmark dataset for fine-tuning or model improvements. However, as pointed out by \cite{balloccu2024leak}, it can be challenging for model owners to identify benchmark data and filter it out to avoid training on it. In case of closed models, the model owner may want to keep models and their weights private. To account for all of the above scenarios, we build an end-to-end system, \tool\footnote{\tool, stands for TRusted UnContaminated Evaluation}, that implements private benchmarking in a variety of trust models (ranging from when some entity such as the dataset owner, model owner or a third party may be trusted, to scenarios where no entity may be trusted and confidential computing or cryptographic solutions must be deployed). We demonstrate the practical feasibility of our system through experimental evaluation. 

Another question that becomes relevant if the benchmark datasets are kept private is: how does one guarantee that it has test data points that are of good quality? Ideally, we want solutions where an auditor is able to check a private benchmark for quality. In this work, we present unique solutions to these problems by introducing advances in confidential computing and cryptographic protocols to the NLP benchmarking community. As far as we are aware, this is the first work that has approached the problem of benchmark data contamination with this perspective.

\section{Challenges with LLM benchmarking today}

\subsection{Lack of access to proprietary benchmarks}

Major developers of LLMs continuously create benchmarks to internally evaluate and iteratively improve their models. Creating such benchmarks requires considerable expertise and time. Such internal benchmarks also hold a competitive advantage to the model owner, as their secret sauce that results in their models outperforming others. Hence model builders in the industry do not always share their benchmarks. Several commercial organizations building applications on top of LLMs also utilize their own internal benchmarks based on user data to evaluate their own models before shipping to production - however, such datasets are not typically made available for research use due to concerns about privacy and sensitivity of the data. If there are mechanisms for competing models to be evaluated on competing benchmarks with neither the model details nor the benchmarks data set being revealed to either party, it will greatly enhance credible benchmarking and faster improvements of models themselves. Access to more benchmarks created by industry will also boost LLM and evaluation research, particularly in the academic community.


\subsection{Data contamination in LLMs}

In the context of Large Language Models (LLMs), contamination of benchmark datasets refers to the presence of test data in either the pre-training or fine-tuning data of an LLM. Such test data can either be present in full or as a subset, and may also include derivatives or transformed versions of the original test data. For instance, certain multilingual benchmarks like XNLI \cite{conneau2018xnli} and MLQA \cite{lewis2020mlqa} are translations of a benchmark originally created in English. It is possible that the original English benchmark may be present in an LLM's data, even if the multilingual version is not. Due to this, despite no contamination of the multilingual dataset, the LLM may still be able to recover an answer from the contaminated data through cross-lingual transfer. \cite{balloccu2024leak} introduce the concept of indirect data leakage, which happens when benchmark data is leaked when researchers evaluate models via APIs or interfaces, where the usage policy states that user interaction data can be used for improving the models. \cite{balloccu2024leak} review several papers on benchmarking and find that 42\% of papers they review may have inadvertently leaked data to GPT-3.5 and GPT-4 due to their access mechanism, which does not guarantee that conversational data will not be used for fine tuning or improving models. According to the study, based on the benchmarks used in the surveyed papers, a total of 4.7M benchmark samples across 263 benchmarks have been leaked in this way. 


Several techniques have been proposed to detect contamination. One of the early heuristic based techniques include checking whether the model under consideration was trained before or after the benchmark was released. While this is an effective way to check for contamination in the original model, it does not account for model updates and indirect data leakage mentioned above. Another heuristic is to check whether the dataset is available as text on the web, or whether it can be only accessed via download. In some cases, datasets are provided in password protected folders, which prevents automated crawlers from ingesting this data. However, there is no guarantee that other copies of these datasets do not exist on the web, or that model builders have not included this data during pretraining or fine-tuning.

Prompting-based techniques detect contamination by asking the model to complete model cards or data points in the dataset \cite{ahuja-etal-2023-mega, golchin2023time}. \cite{deng2023investigating} prompt models to fill in a masked answer, which is the incorrect answer in a multiple choice setup, and also prompt the model to predict an unlikely word in a test example. \cite{li2023task} use training data inspection, task example extraction, and a membership inference attack to suggest that LLM improvements can be explained by test data contamination. \cite{golchin2023data} propose a technique in which they generate perturbations of the benchmark datasets while preserving meaning. They provide the perturbations with the original text as options for the model to pick from, the idea being that a model contaminated with the benchmark will pick the original data point more often that the other data points even though they share the same semantics. While this approach has been somewhat effective in detecting contamination, it may not perform well in scenarios where the perturbations, generated by an LLM, are of inferior quality and do not preserve semantics, such as multilingual settings \cite{ahuja2023megaverse} where the LLM does not perform this task well in non-English languages.

While it is desirable for many reasons to open-source LLMs, it is likely that some models will not be made available to the public. Additionally, many open-source models do not disclose information about the training dataset, making it challenging to detect contamination. Moreover, several open-source models are constructed on top of existing open-source models which may not reveal training dataset details (e.g., fine-tuned versions of the Llama \cite{touvron2023llama} models), meaning that contamination continues to be a significant challenge. Overall, while it is an important endeavour to be able to detect contamination in models, existing techniques are not fool-proof, and hence, it is critical to prevent contamination in the first place.
\section{Model Evaluation on Private Benchmarks}

To prevent contamination, we propose a solution in which the benchmark dataset is kept private, which would by design, ensure that the model has never seen the benchmark dataset. However, doing so is not straightforward and faces the following two challenges: 1) model publishers of proprietary LLMs typically wish to keep their models (or at least model weights) private; how do you then evaluate models on private benchmarks while keeping the model weights private? 2) If the benchmarks are kept private, how can we audit the benchmark to ensure that it is of sufficiently high quality? 

We describe various scenarios in which we trust various actors in the process - the model owner, dataset owner, a trusted third party etc. We articulate different methods in which private benchmarking can be achieved in all these scenarios highlighting their pros and cons. To answer the first question (evaluating models on private benchmarks keeping model weights private), we discuss existing solutions to the problem in Section~\ref{subsec:existingsolutions} and present recent technological advancements that provide better solutions to this in Section~\ref{subsec:newsolutions}. We demonstrate the practical feasibility of our approach through an end-to-end system described and evaluated in Section~\ref{sec:system}. In Section~\ref{sec:auditing}, we answer the second question and discuss how private benchmarks can be audited.

Let us fix some notation. Let the model architecture (function description) without the model weights be denoted by $M$. Let the weights of the model be denoted by $w$ and let the inputs in the benchmark dataset be denoted by $\vec{x}= x_1, \cdots, x_t$. These inputs may also be labelled with labels $\vec{l}= l_1, \cdots, l_t$. Typically the objective of an evaluation would be to evaluate every input $x_i$ on the model $M$ with weights $w$ and check if the output of this matches the label $l_i$. That is, we want to output the bit $z_i$ which is $1$ when $M(w,x_i) = y_i$ and $0$ otherwise. These results can be then collated to compute, for example, the accuracy of the model $\mathsf{Acc}_{M,w,\vec{x},\vec{l}}=(\sum_{i=1}^t z_i)/t$. Other functions can also be computed on the model's outputs. 


\subsection{Existing Solutions}\label{subsec:existingsolutions}

\subsubsection{Model as an API (Trusted model owner)} 
\begin{figure}
\begin{minipage}{0.5\textwidth}
    \includegraphics[scale=0.26]{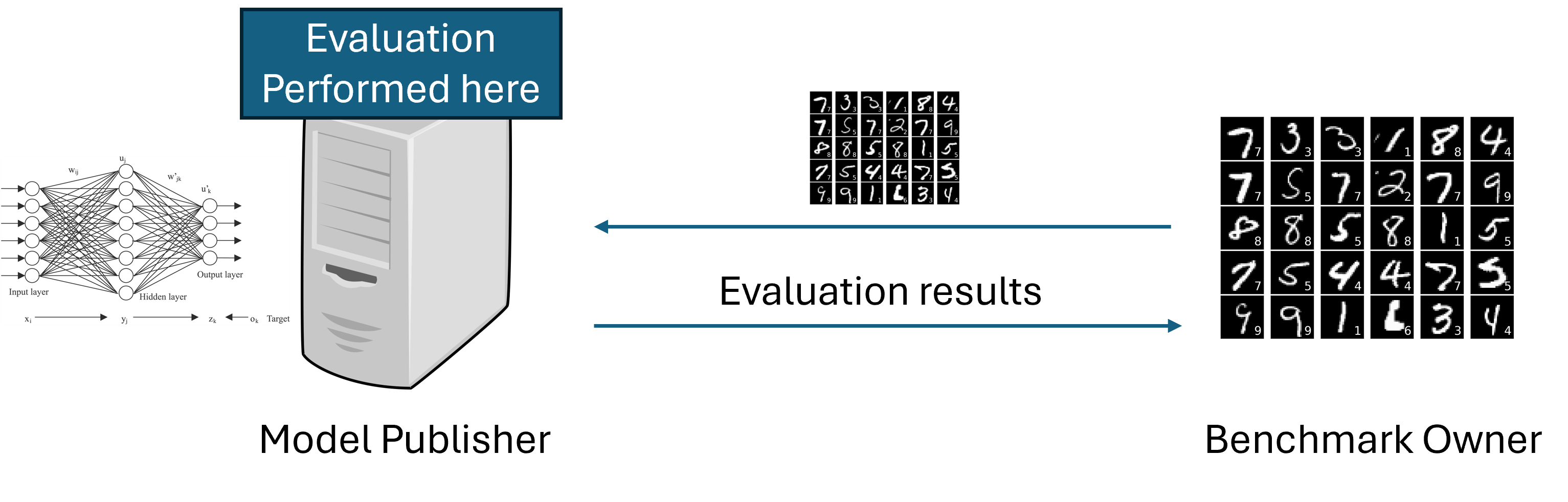}
    \captionsetup{singlelinecheck=false,justification=centering}
    \caption{Model as an API \\(Trusted Model Owner)}
    \label{fig:llmeval1}
    \end{minipage}
    \begin{minipage}{0.5\textwidth}
    \includegraphics[scale=0.26]{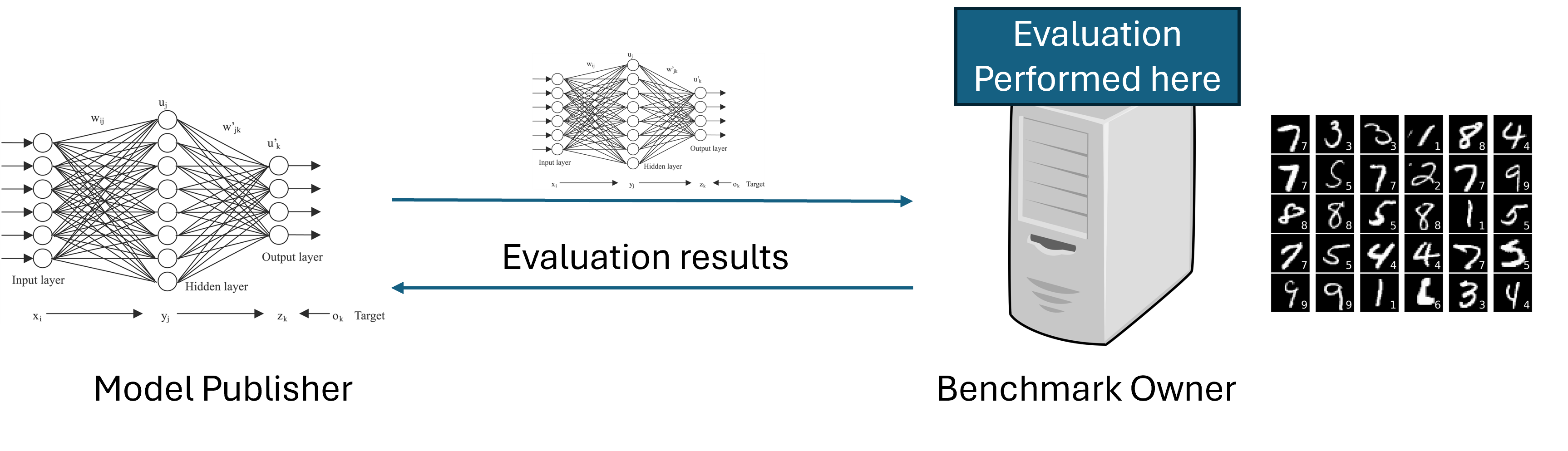}
    \captionsetup{singlelinecheck=false,justification=centering}
    \caption{Open-source Models \\(Trusted dataset owner)}
    \label{fig:llmeval2}
    \end{minipage}
\end{figure}

If the model is only available as an API (e.g., in the case of proprietary models such as the OpenAI models), then one can only query these models on benchmark inputs and compute the accuracy of the returned results. Since the model publisher (e.g., OpenAI) will get to see the queries being made to the model (which would be the benchmark dataset in this case), in this scenario, the model publisher would have to be trusted to not train/contaminate the models on the benchmarks. While in this solution, the model architecture as well as weights are kept private, the benchmark inputs are revealed to the model publisher and one must rely on the publisher ensuring that the model has not been contaminated by the benchmark. The solution architecture for this case is illustrated in Figure~\ref{fig:llmeval1}.

\subsubsection{Open-source Models (Trusted dataset owner)} In this scenario, the model is open-source (e.g., Llama2 \cite{touvron2023llama}) but the benchmark dataset is kept private by the benchmark owner (e.g., Scale~\cite{scale}). In this case, the model evaluation can be done on the private benchmark dataset and the results of the evaluation can be made available. The owner of the benchmark dataset performs the evaluation by hosting the model themselves and must be trusted to perform the evaluation correctly. This guarantees that the benchmark dataset does not flow back into the model for making model improvements. The solution architecture for this case is illustrated in Figure~\ref{fig:llmeval2}.

 
\subsubsection{Trusted Evaluation of models by a Third Party (Trusted third party)} In this scenario, a trusted third party can host a computation environment that is trusted by all entities. This environment can receive both the model as well as the benchmark dataset and the evaluation of the model can be done in the clear within this environment. Hence, one relies on the trusted third party to keep both the model architecture/weights private as well as to ensure that the model has not been contaminated by the benchmark. The solution architecture for this case is illustrated in Figure~\ref{fig:llmeval3}.

\subsection{Proposed New Solutions}\label{subsec:newsolutions}
We propose two approaches to private benchmarking based on  different technologies; see~\cite{queue} for a discussion of tradeoffs among them. 
\subsubsection{Confidential Computation of models on the platform (Hub and Spoke)} 
\begin{figure}
\begin{minipage}{0.5\textwidth}
    \includegraphics[width=\textwidth]{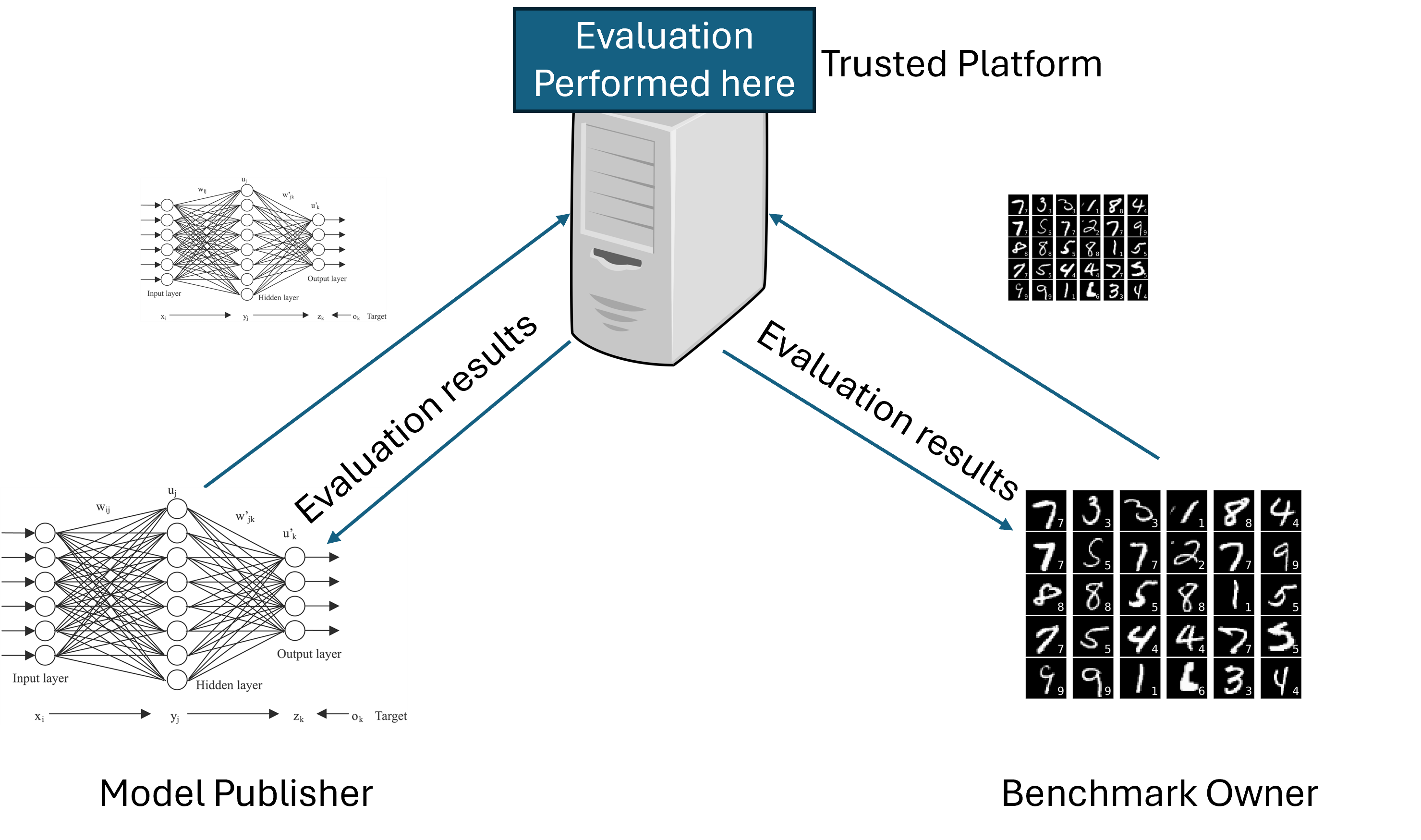}
    \captionsetup{singlelinecheck=false,justification=centering}
    \caption{Trusted Evaluation by a Third Party \\(Trusted Third Party)}
    \label{fig:llmeval3}
\end{minipage}
   \begin{minipage}{0.5\textwidth}
    \includegraphics[width=\textwidth,height=4.8cm]{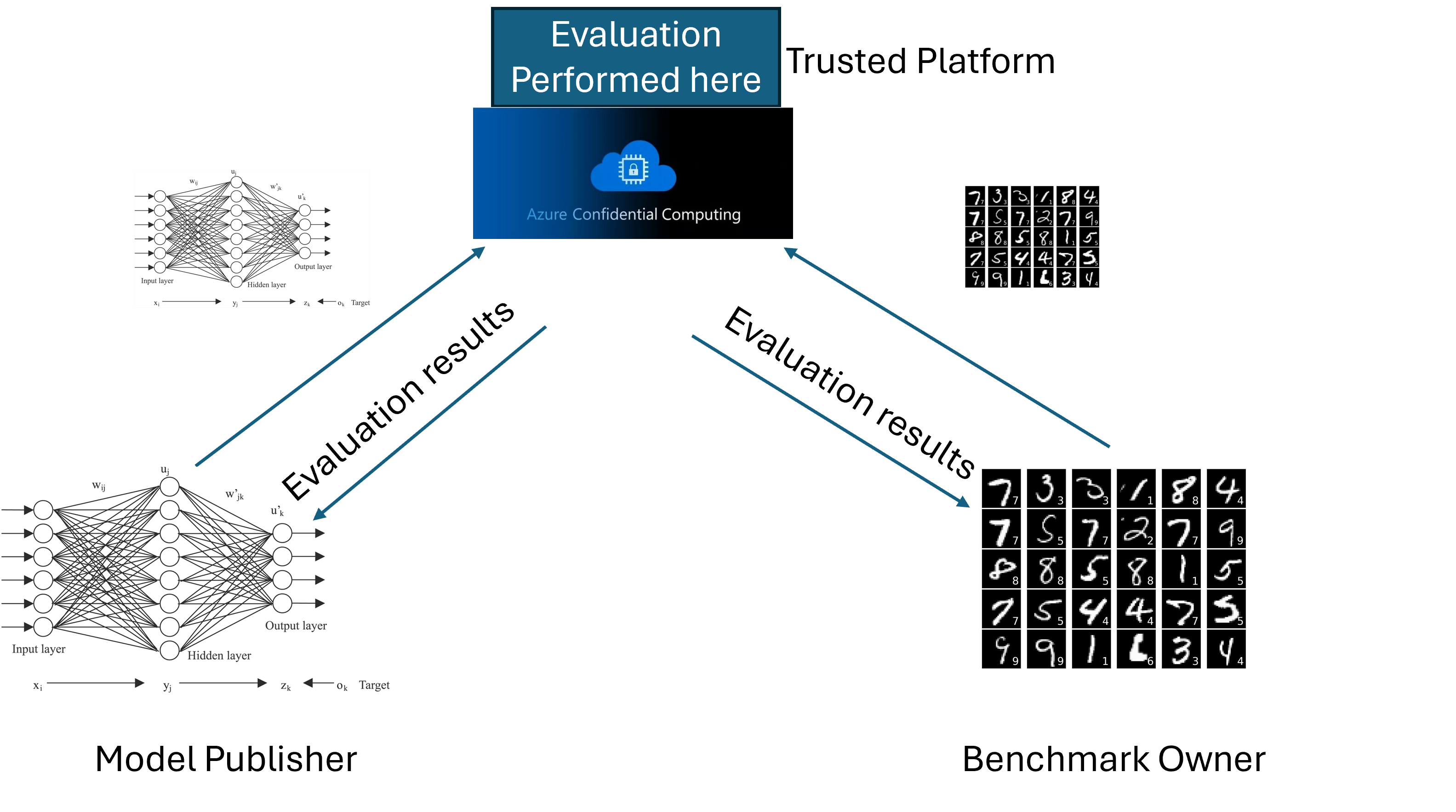}
    \captionsetup{singlelinecheck=false,justification=centering}
    \caption{Confidential Computation \\(Hub and Spoke)}
    \label{fig:llmeval4}
   \end{minipage} 
\end{figure}

In this scenario, the computation environment hosted will be a trusted execution environment (TEE) such as Azure Confidential Computing \footnote{https://azure.microsoft.com/en-us/solutions/confidential-compute/}, Google Confidential Computing \footnote{https://cloud.google.com/security/products/confidential-computing} or AWS Nitro System \footnote{https://aws.amazon.com/ec2/nitro/}. This TEE cannot be accessed by any entity (including the service hosting it). The TEE can be a confidential Virtual Machine (VM), a hardware enclave (such as Intel SGX) or Hopper and Blackwell GPUs by NVIDIA in confidential mode. The model publisher and benchmark dataset owner can encrypt the model and benchmark respectively under the public key of the confidential environment. Within the TEE, the model and the benchmark can be decrypted, evaluation can be performed, and the results of the evaluation can be made available.
This solution has the advantage that both the model weights as well as the model architecture can be kept hidden from the benchmark owner. 
The solution architecture for this case is illustrated in Figure~\ref{fig:llmeval4}. While confidential computing had scalability issues in the past because of the small size of the memory in the TEE, this is no longer an issue in the latest systems based on confidential GPUs~\cite{hoppercc}.  
 
\subsubsection{EzPC technology to evaluate models on benchmarks (P2P)} EzPC~\cite{ezpc,cryptflow2,orca,sigma,ezpc-code} is a cryptographically secure system built on the basis of the seminal cryptographic technology of secure multi-party computation~\cite{gmw,yaogc}. It allows multiple entities to bring in private data and enable them to perform computations on the data without trusting any entity or each other in the computation. This is done through an interactive protocol in which parties perform computations as well as interact and exchange (seemingly random looking) messages with the other party. The security guarantees provided are cryptographic (i.e. unless cryptography is broken, e.g. AES-128 is broken, no entity can learn any information other than the results of the computation, which in this case would be the evaluation results). 

\begin{figure*}
    \centering
    \includegraphics[scale=0.35]{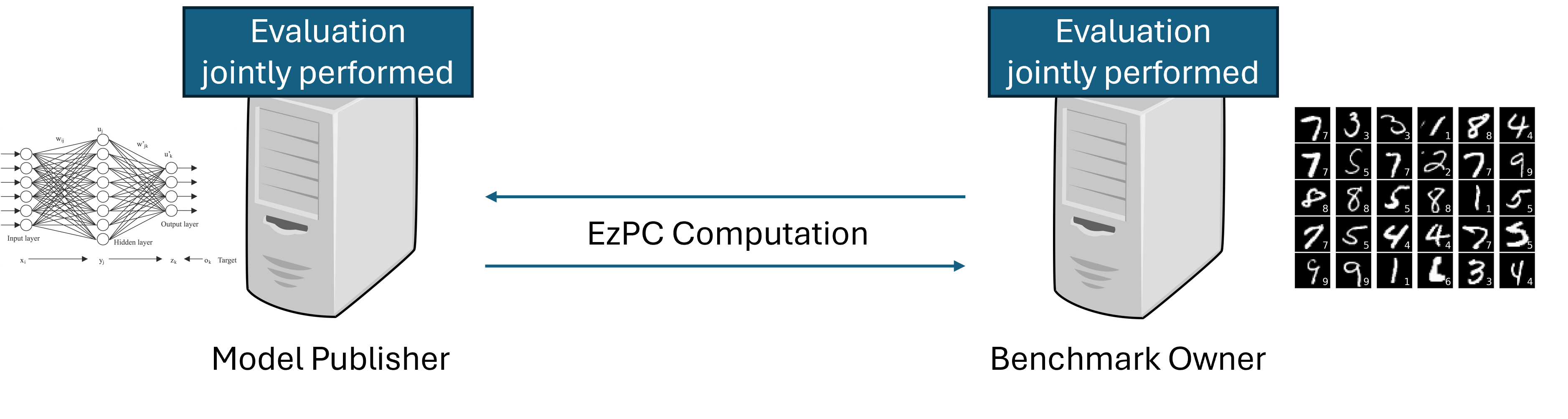}
    \caption{EzPC based on Secure Multi-Party Computation (P2P)}
    \label{fig:llmeval5}
\end{figure*}

EzPC technology enables secure computation of LLMs and provides support for execution on CPUs as well as GPUs. It has also been developed and tested in the context of model evaluation in the healthcare domain (see~\cite{secureeval}). Secure multi-party computation based solutions such as EzPC assume that the model architecture is known to all participants. While the most efficient systems today (such as the EzPC system) typically assume that the participants execute the protocol code given to them faithfully (called {\em semi-honest behavior} in cryptography), solutions also exist that avoid such an assumptions ~\cite{mpspdz,mpspdz-code}. The solution architecture for this case is illustrated in Figure~\ref{fig:llmeval5}. While solutions based on cryptographic secure multiparty computation suffered from astronomical overheads in the past, the latency overheads of EzPC are tractable~\cite{sigma}. 

\section{System and Evaluation}\label{sec:system}
\subsection{Implementation} We have implemented an open-source platform, called \tool, to help model owners and dataset owners to collaborate and create public leaderboards.
For evaluation, the platform supports all five benchmarking options: trusted model owner (Figure~\ref{fig:llmeval1}), trusted dataset owner (Figure~\ref{fig:llmeval2}), trusted third party (Figure~\ref{fig:llmeval3}), trusted execution environment (Figure~\ref{fig:llmeval4}), and cryptographic secure multiparty computation with EzPC (Figure~\ref{fig:llmeval5}).
The platform is implemented in 5k lines of python using Django as the backend.

A model owner can raise a request to evaluate the available private datasets using any of the five benchmarking methods. Conversely, this mandates that both the model owner and dataset owner agree to run the evaluation.
For each type of evaluation submission, the platform will generate a unique ephemeral certificate, signed by the platform’s Certificate Authority, for that submission only. This certificate is then used to ensure encrypted communication using TLSv1.3 and authenticate parties to each other in TEE, TTP (third party), and EzPC (P2P). After the evaluation is complete, the results are pushed to the platform's public leaderboard.

The platform is publicly available at \url{github.com/microsoft/private-benchmarking} under MIT license.

\subsection{Evaluation}\label{eval}

\begin{table}
    \centering
    \renewcommand{\arraystretch}{1.5} 
    \caption{Evaluating models on datasets using a Trusted Third Party (TTP), a Trusted Execution Environment (TEE), and with EzPC. We report the inference time per sample, the number of samples in the test set, and the total communication of evaluating the entire test set.}
    \hfill \break
    \label{tab:communication_time}
    \begin{tabular}{l|rrrr}
    \hline
         \textbf{Model + Dataset} & \textbf{Type} & \textbf{Time per sample (s)} & \textbf{\#samples} & \textbf{Total Communication (GB)} \\ \hline
        \textbf{Bert-Base~\cite{devlin2018bert}} & TTP & 0.00137  & ~ & 0.40 \\ 
        \textbf{on} & TEE & 0.00161 & 1,821 & 0.40 \\ 
        \textbf{SST2~\cite{socher-etal-2013-recursive}} & EzPC & 1.322 & ~ & 1,442.13  \\ \cline{1-5}
        \textbf{Airavata~\cite{gala2024airavata}} & TTP & 0.00832 & ~ & 25.60  \\ 
        \textbf{on} & TEE & 0.00845 & 5,010 & 25.60  \\ 
        \textbf{Indicxnli~\cite{aggarwal-etal-2022-indicxnli}} & EzPC & 12.157 & ~ & 75,711.03  \\ \cline{1-5}
        \textbf{Llama2-7B~\cite{touvron2023llama}} & TTP & 0.05298 & ~ & 25.53  \\ 
        \textbf{on} & TEE & 0.05354 & 5153 & 25.53  \\ 
        \textbf{Lambada~\cite{lambada_dataset}} & EzPC & 12.269 & ~ & 1,98,996.58  \\ \cline{1-5}
        \textbf{VGG16~\cite{simonyan2015deep}} & TTP & 0.00055 & ~ & 0.69  \\ 
        \textbf{on} & TEE & 0.00058 & 10,000 & 0.69  \\ 
        \textbf{ImageNet-1k~\cite{russakovsky2015imagenet}} & EzPC & 0.329 & ~ & 3,166.97  \\ \hline
    \end{tabular}
\end{table}

We evaluate the time and communication of various private benchmarking options on publicly available models and datasets in Table~\ref{tab:communication_time}. We find that compared to the existing baseline solutions (Section~\ref{subsec:existingsolutions}), the overheads of confidential computing (using confidential Hopper GPUs) are negligible and that of EzPC (using Sigma~\cite{sigma}) are tractable.

\paragraph{Evaluation setup.} We evaluate on Azure virtual machines (VMs) where each VM has an Nvidia's Hopper H100 GPU, 40 CPU cores, and 320 GB RAM.  To benchmark using a trusted third party (TTP), both the model and the dataset are loaded into a single VM which runs the evaluation on the GPU. For evaluation with trusted execution environment, the GPU is put in the confidential mode. For evaluation with EzPC, we use two such VMs that are connected via a computer network with 24Gbps bandwidth.

\paragraph{Performance.} The baseline TTP time for the language models is obtained by running ten samples with 100 tokens each and taking the mean of the runtimes. 
For the vision model VGG16, we run it on ten $224\times 224\times 3$ images and take the mean. We follow the same process for TEEs and the running time increases slightly. The running time of cryptographic computation with EzPC is higher but tractable. The variations in runtime of the different mechanisms are negligible.   

Apart from time, another metric of interest is communication. The TTP and TEE options must communicate the model and the dataset to the evaluating VM. Subsequently, no communication is required during the evaluation process. With EzPC, the VM belonging to the model owner
and the VM belonging to the dataset owner must communicate encrypted data for each inference. While the communication of EzPC is much higher, cryptographic solutions are becoming more communication frugal with each passing year~\cite{ezpc,cryptflow2,orca,sigma}. Note that the time per sample reported for EzPC in Table~\ref{tab:communication_time} includes the time taken in communicating the encryptions as well.

\section{Auditing Datasets}\label{sec:auditing}

One of the crucial aspects of benchmark creation is ensuring that the benchmark is of high quality and can reliably test model capabilities of interest. To create a representative evaluation dataset, several factors must be taken into consideration, including the source of the data, data selection and filtering practices, as well as annotator demographics and language background \cite{dogruoz-etal-2023-representativeness}. Neglecting these factors can result in low-quality datasets that provide misleading results. Therefore, it is crucial for us to be able to audit benchmarks to ensure that they are of sufficiently high quality. In the case of open benchmarks, this is relatively easy to do. Similarly, owners of large datasets can randomly partition their datasets into training sets which are released publicly and keep small test sets private~\cite{gaia}.

In the case of private benchmarks, we might want to enable an auditor to test various properties of the benchmark (e.g. to test the quality of the benchmark). As long as this test can be expressed as a function $T$ that operates on the model and inputs, it can be computed in a similar manner to performing model evaluation. Re-using notation established in the earlier section, if the audit function can be expressed as a function $\mathsf{Audit}(M,w,\vec{x},\vec{l})$, where $M$ is the model architecture, $w$ are the weights of the model, $\vec{x}=x_1, \cdots, x_t$ is a set of input points to the model and $\vec{l} = l_1, \cdots, l_t$ are labels to these inputs (if needed), then one can also compute the above function using the same techniques described in the previous section. In this way, only the audit functions output would be revealed to the auditor (which can even be as fine grained as simply whether the model passed the audit or not). 

\paragraph{Lightweight Auditing.} We now describe different methods using which auditors can audit private benchmarks. These methods have their pros and cons which we discuss below. 

{\em Honest Benchmark Owner.} In this, we assume that the benchmark owner is an honest entity. That is, the benchmark owner curates a benchmark to be presented to the auditor. While this benchmark may not be of a high quality, the owner will not act maliciously to fool an auditor into believing otherwise. In this case, the following method can be used to audit a benchmark. As before, suppose the benchmark has data points $\vec{x}= x_1, \cdots, x_t$ with labels $\vec{l}= l_1, \cdots, l_t$. Now, the auditor samples $\kappa$ indices at random from $\{1, 2, \cdots, t\}$ and sends these indices $\mathcal{I} = \{i_1, \cdots, i_\kappa\}$ to the benchmark owner. The benchmark owner provides the auditor with $\{(x_{i_1}, l_{i_1}),(x_{i_1}, l_{i_2}), \cdots, (x_{i_\kappa}, l_{i_\kappa})\}$. We also assume that the test $T$ can be performed individually on each benchmark data point to determine its quality (i.e., $T(x_{i},l_i)$ returns $1$ or $0$ depending on whether the benchmark is of good quality or not). Now, the auditor performs the test $T$ on each of these $\kappa$ points and declares the benchmark to be of a good quality if $T(x_{i},l_i) = 1$ for all $i \in \mathcal{I}$. One can then easily show a theorem of the following form: ``If the auditor declares the benchmark to be of a good quality, then at least $\alpha \%$ of the dataset is of good quality, except with probability $\approx (\alpha/100)^\kappa$". For example, if we wanted a guarantee that $95\%$ of the dataset is of a good quality, then by checking $\kappa = 100$ points at random, we can ascertain that this is indeed the case, except with probability $0.006$. Of course, this solution will reveal 100 data points to an auditor and is only applicable when this is acceptable to the benchmark owner.

{\em Somewhat Honest Benchmark Owner.} One can also design a solution when we do not wish to assume that the benchmark owner is fully honest as in the above solution. Suppose the benchmark owner curates a benchmark to be presented to the auditor, but the owner may also maliciously try to fool the auditor into believing that this benchmark if of high quality. One can provide a solution in this case as follows. As above, we assume that the benchmark has data points $\vec{x}= x_1, \cdots, x_t$ with labels $\vec{l}= l_1, \cdots, l_t$ and as above, we assume that the test $T$ can be performed individually on each benchmark data point to determine its quality (i.e., $T(x_{i},l_i)$ returns $1$ or $0$ depending on whether the benchmark is of good quality or not). In this solution, the benchmark owner will provide a cryptographic commitment~\cite{goldreichbook,merklecommitment} to each of the $t$ data points (a cryptographic commitment to a value $v$ can be thought of as providing a locked box that contains the value $v$. The commitment (locked box) hides the value $v$ and can be opened later (with an opening or a key). The value inside a locked box cannot be changed (called binding) once sent to the recipient). Once the benchmark owner provides a cryptographic commitment $C_i$ to every $(x_i,l_i), i \in \{1, \cdots, t\}$, the auditor then does as in the previous solution. That is, the auditor samples $\kappa$ indices at random from $\{1, 2, \cdots, t\}$ and sends these indices $\mathcal{I} = \{i_1, \cdots, i_\kappa\}$ to the benchmark owner. The benchmark owner will now provide the openings (keys) to each of the $\kappa$ commitments (i.e., for $C_{i_1}, \cdots, C_{i_\kappa}$). The auditor will open these commitments, learn the $\kappa$ data points and perform the test $T$ on these points. The guarantees provided are as before - i.e., one can easily show ``If the auditor declares the benchmark to be of a good quality, then at least $\alpha \%$ of the dataset is of good quality, except with probability $\approx (\alpha/100)^\kappa$".

It is to be noted here that while the above solution prevents a malicious benchmark owner from presenting a poor dataset to the auditor, the malicious owner can still present a good dataset to the auditor but use a poor dataset in the evaluation phase with a model owner. In order to further ensure that the same dataset is used in both auditing and evaluation, the benchmark owner must provide a cryptographic commitment to the benchmark that is made available to both model owners as well as auditors. Then, during the model evaluation process, the benchmark owner must prove (using a cryptographic zero-knowledge proof) that the benchmark that was committed to was indeed the same one used in the evaluation process. This mechanism is easier in the case of TEE based solutions as zero-knowledge proofs are unnecessary and the TEE can simply receive the commitment and the opening of the commitment to the benchmark and use that benchmark in the evaluation process.

\section{Conclusion}

In this paper, we highlight the problems caused by benchmark contamination in LLM training data. Given that benchmarking is the de-facto standard for evaluating LLMs and applications built on top of them, it is critical that we address this problem. There have been several attempts at trying to detect contamination in LLMs post-hoc, however, none of them are fool-proof, and hence it is important that we try and prevent contamination in the first place.

We present a unique inter-disciplinary solution to this problem - private benchmarking, in which the benchmark datasets are kept private from the model, using advances in secure computation and cryptography. We present solutions to several trust models, describe their pros and cons, build an end-to-end platform for such private benchmarking and compare the various solutions on performance aspects.
Another possibility is that of startups offering private benchmarking  as a service using the provably secure techniques described in this paper (such as in Figure 5) for keeping the model weights and the benchmark data private. This is similar to the creation of companies at the start of the RSA security  certification process in the early 90s and the emergence of large companies like Verisign. Another example is that of the Truste seal \footnote{https://www.trustsignals.com/blog/history-of-truste} used to signal privacy practices of e-commerce websites in the early 2000s.

High-quality benchmarks provide useful insights for improvement, while low-quality benchmarks can provide misleading results. It is important to be able to audit the quality of benchmarks even in the private benchmarking setting. We present solutions to this problem with assumptions about the honesty of the benchmark owner and show that it is possible to audit private benchmarks.

In conclusion, we hope that this paper spurs interest in the problem of preventing benchmark contamination and enabling the sharing of proprietary benchmarks by using novel solutions such as private benchmarking. We also hope that our paper encourages the NLP community to collaborate with other disciplines for solving some of the hard challenges we face today.

\section{Limitations}

In this paper, we present techniques that can potentially be used for private benchmarking to solve the issue of dataset contamination. Although we present solutions for auditing datasets that ensure that the benchmark remains private, it is possible that we are not able to identify issues with benchmarks using these auditing techniques. Additionally, keeping datasets private may have other unintended consequences, such as increasing barriers to access such datasets, in comparison to open source datasets. 

\section*{Acknowledgements}
We thank Kapil Vaswani and Krishnaprasad Hande for providing us access to confidential GPUs. We thank Neha Jawalkar and Kanav Gupta for help with EzPC.

\nocite{Ando2005,andrew2007scalable,rasooli-tetrault-2015}


\printbibliography

\end{document}